\documentclass[12pt]{article}

\usepackage{bm}
\usepackage{amssymb}
\usepackage{enumerate}

\usepackage{graphicx} 

\newcommand\ee{ e^+ e^-}
\newcommand\zdecay{ Z' \to e^+ e^-}
\newcommand\znudecay{ Z' \to \nu \nu }

\begin{document}


\title{ Probing the muon $g_\mu -2$ anomaly,  $L_{\mu} - L_{\tau}$ gauge boson   and Dark Matter in dark photon experiments}               

\author{ S.N.~Gninenko$^{1}$ and  N.V.~Krasnikov$^{1,2}$   
\\
$^{1}$ Institute for Nuclear Research, 117312 Moscow 
\\
$^{2}$ Joint Institute for Nuclear Research,141980 Dubna}





\maketitle

\begin{abstract}
In the $L_{\mu} - L_{\tau}$ model the 3.6 $\sigma$ discrepancy between the predicted and  measured values of the 
anomalous  magnetic moment of positive muons can be explained by  
the existence of a new dark boson $Z'$ with a mass in the sub-GeV range, which is coupled at tree level predominantly to the second and third lepton generations.  However, 
at the one-loop level  the   $Z'$ coupling  to electrons or quarks can be   induced via the $\gamma -Z'$ kinetic mixing, which  is generated  through the loop
 involving the muon and tau lepton. 
This  loophole has important experimental consequences since it opens up  new possibilities, in particular  for the complementary searches of the $Z'$ in the ongoing NA64 and incoming   dark photon experiments with high-energy electrons. An extension of the model able to explain relic Dark Matter density is also discussed.

\end{abstract}


\newpage
At present there are several  signals that    new physics beyond the standard model (SM) exists. The most striking is the observation of
Dark Matter (DM).  Among the many  models  of DM, for a review, see e.g. \cite{Universe0} - \cite{Universe3}, those  that  motivate  
the existence of  light vector(scalar) bosons with a mass $m_d \leq O(1)~GeV$  are  rather popular now  \cite{lightdark,lightdark1}. 
The main idea  is that  in addition to gravity a new  interaction between  visible and dark sector exists which is 
mediated by this gauge boson  \cite{lightdark1}. 
\par Anther possible hint in favour of new physics is the   muon $g_{\mu}-2$ anomaly, which is 
the 3.6 $\sigma$ discrepancy between the 
experimental values \cite{exp1,pdg}  and the SM  predictions \cite{th1,th2,th3,th4}  for  the anomalous magnetic moment of the muon. 

Among several  extensions of the SM explaining the anomaly, the   models predicting the existence  
of  a weak leptonic force  mediated by a sub-GeV gauge boson $Z'$ that couples predominantly  to the difference between the muon and tau lepton numbers, $L_{\mu} - L_{\tau}$,  
are of general interest. The abelian symmetry $L_{\mu} - L_{\tau}$ is an anomaly-free global symmetry
within the SM \cite{foot1, foot2, foot3}. Breaking $L_{\mu} - L_{\tau}$  is  crucial  for the appearance  of a new relatively light, with 
a mass $m_{Z'} \leq 1~GeV$,  vector boson ($Z'$) wich couples very weakly to  muon and tau-lepton with the 
coupling constant $\alpha_\mu \sim O(10^{-8})$ \cite{vecmuon1}- \cite{vecmuon8}.
 The $Z'$ interaction  with $L_{\mu} - L_{\tau}$ vector current   given by 
\begin{equation}
L_{Z'} = e_{\mu}Z'_{\nu}[\bar{\mu}\gamma^{\nu}\mu - \bar{\tau}\gamma^{\nu}\tau  +  \bar{\nu_{\mu}}\gamma^{\nu}\nu_{\mu} 
 - \bar{\nu_{\tau}}\gamma^{\nu}\nu_{\tau} ]
\label{lagrangian}
\end{equation} 
leads to  additional contribution to the  muon 
anomalous magnetic moment \cite{muonmoment}     
\begin{equation}
\delta a  = \frac{\alpha_{\mu}}{2\pi} F(\frac{m_{Z'}}{m_{\mu}}) \,,
\end{equation}
where
\begin{equation}
F(x) = \int^1_0 dz \frac{[2z(1-z)^2]}{[(1-z)^2 + x^2z]} \,
\end{equation}
and $\alpha_{\mu} = \frac{e_{\mu}^2}{4\pi}$. 
  The use of the formulae  (2,3) allows to determine the coupling constant $\alpha_{\mu}$ which 
explains the value of the  $g_{\mu} - 2$ anomaly \cite{vecmuon1} - \cite{vecmuon8} 
and it does not contradict to existing  experimental bounds for $m_{Z'} \leq 2 m_{\mu}$ \cite{vecmuon8}. 
Namely,  for $ m_{Z'} \ll m_{\mu}$  \cite{muonmoment}
\begin{equation}
\alpha_{\mu} = (1.8 \pm 0.5) \times 10^{-8}\\.
\label{alpha1}
\end{equation}
For another limiting case  $ m_{Z'} \gg m_{\mu}$ the $\alpha_{\mu}$ is 
\begin{equation}
\alpha_{\mu} = (2.7 \pm 0.7) \times 10^{-8} \times \frac{m^2_{Z'}}{m^2_{\mu}}   \\.
\label{alpha2}
\end{equation}
In addition to the case of the $g_{\mu}-2 $  anomaly, there are also other implications  of  $ Z'$  \cite{vecmuon1}- \cite{vecmuon8}. For example, 
in neutrino sector,  the   $L_{\mu} - L_{\tau}$  model can  provide a natural explanation of a zeroth-order approximation for neutrino mixing with a quasi-degenerate mass  spectrum  predicting  a  maximal  atmospheric  and  vanishing reactor neutrino mixing angle \cite{nu1, nu2, nu3},  small masses of neutrinos and its oscillations by extending the model with the left-right gauge symmetry \cite{dev}, the $R_K$ puzzle in LHCb data and the $g_{\mu} - 2 $ anomaly  can be simultaneously explained with the $\simeq 10$ MeV $Z'$   which also induces  the nonstandard matter interactions (NSI) of neutrinos  \cite{datta}.  The later could also provide LMA-Dark solution to solar anomaly,  which also requires NSI \cite{farzan}. 
 Recently,  it has been pointed out that specific features of cosmic neutrino spectrum reported by the IceCube Collaboration can be explained by 
  a mass of the MeV scale \cite{takeshi0, takeshi1}, which can be of interest for the search at Belle II \cite{takeshi2}. 
Moreover, below we show that the $L_\mu - L_\tau$  model with a $\simeq$ 10 MeV $Z'$  boson interacting 
with a light DM  can  also explain the observed relic DM  density. All these solutions employ  a  SM extension with a gauge $L_\mu-L_\tau$  model. 
  
It is generally assumed that searches  for the  $L_\mu - L_\tau$  gauge boson are difficult as it   couples only to the muon and  tau lepton family.  The relevant  bounds  can be extracted form the measurements of  the neutrino trident  production   $\nu_\mu N \to \nu_\mu \mu^+ \mu^- N$  \cite{CHARM, CCFR}, from  the search for a muonic dark force at BABAR \cite{BABAR1},   and from the data of  the Borexino experiment \cite{BOREXINO}. 
Currently,   the allowed $Z'$ mass region  for the explanation  of the $g_{\mu} -2$  anomaly is  constrained 
to $ m_{Z'} \lesssim  400$ MeV from by the neutrino trident  production  \cite{nutri, ccfr}, while   the BABAR search excluded masses  $ m_{Z'} \gtrsim  200$ MeV.  Besides this,  if the  $Z'$  is light  it would increase the number of relativistic degrees 
of freedom that spoils the success of the standard Big Bang nucleosynthesis (BBN) predictions.  This leads to the 
 lower bound $m_{Z'} \geq 1$~MeV \cite{BBN}. Moreover there is more restrictive 
bound $m_{Z'} \geq (3-5)~MeV$ \cite{BBN1} based on the fact that relatively light $Z^`$ 
may indirectly contribute to the number of effective 
relativistic degrees of freedom $N_{eff}$ through the raise of the 
temperature $\nu_{\mu}$ and $\nu_{\tau}$. From the requirement $\Delta N_{eff} < 0.7$ 
more restrictive bound   $m_{Z^`} \geq 5~MeV$ arises \cite{BBN1}.     
To search for the $Z'$ in the still unconstrained mass region $5\lesssim  m_{Z'}  \lesssim   200$ MeV   is challenging  as 
the $Z^`$  dominant decay $Z'\to \nu \bar{\nu}$ is invisible. 
 The direct search for such  $Z'$  by  using the reaction  $\mu Z \rightarrow \mu Z Z'; Z'\to invisible$  of the $Z'$ production in high-energy muon scattering off heavy nuclei at the  CERN SPS was  proposed  in Ref. \cite{gkm}.   The experiment  is  expected to improve the sensitivity to the coupling $\alpha_\mu$ by 
a few orders of magnitude and fully cover the parameter region  referred with Eqs. (4) and (5).

\par Let us now  discuss the mixing between the $Z'$ and ordinary photons.  An account of one-loop 
diagrams, which is in our case propagator diagrams with virtual $\mu$- and $\tau$-leptons 
in the loop,  leads to nonzero $\gamma -Z'$ kinetic mixing $\frac{\epsilon}{2}F^{\mu\nu}Z^{\mu\nu}$ 
where $\epsilon$ is the finite  mixing strength given by \cite{Holdom} 
\begin{equation}
\epsilon = \frac{8}{3} \frac{ee_{\mu}}{16\pi^2}{\rm ln}(\frac{m_{\tau}}{m_{\mu}}) = 1.43 \cdot 10^{-2} \cdot e_{\mu} \,.
\label{mixstr}
\end{equation}  
Here $e$ is the electron charge  and $m_\mu, ~ m_\tau$ are the muon and tau-lepton masses respectively. 
It should be stressed that we assume 
 that possible tree level mixing  $\frac{\epsilon_{tree}}{2}F^{\mu\nu}Z^{\mu\nu} $ 
is absent or much smaller than one-loop mixing $\frac{\epsilon}{2}F^{\mu\nu}Z^{\mu\nu} $. To be precise, we 
assume that there is no essential cancellation between tree level and one loop mixing terms $|\epsilon_{tree} + \epsilon | \geq  |\epsilon|$ . 
For $m_{Z'} \ll m_{\mu}$ the value $e_{\mu} = (4.75 \pm 0.8) \cdot 10^{-4}$ from  Eq. (\ref{alpha1})  leads to the prediction of the corresponding mixing value   
\begin{equation}
\epsilon = (6.8 \pm 1.1) \cdot 10^{-6}
\label{loop}
\end{equation}
Thus, one can see that  the $Z'$  interaction  with the $L_{\mu} - L_{\tau}$ current induces 
at one-loop level the $\gamma - Z' $ mixing of $Z'$  with ordinary photon 
which allows to probe $Z'$  not only in muon or tau induced reactions but also with intense electron beams. In particular, this loophole  opens up  the possibility 
of  searching  the new weak leptonic force mediated by the $Z'$  in experiments looking for dark photons ($A'$).
 \par The fact that  
 the $\gamma-Z'$ mixing of Eq.(\ref{loop}) is at an experimentally interesting  level  is very exciting. We point out further  that a
 new intriguing possibilities for the complementary searches of the $Z'$ in the currently ongoing 
experiment  NA64 \cite{na64-1,na64-2} exists. Indeed, the NA64 aimed at the direct search for invisible decay of sub-GeV dark photons in the reaction 
$e^- + Z \to e^- + Z + A'; ~ A' \to invisible$  of high energy electron scattering off heavy nuclei \cite{sngna64, NA64exp}. The experimental signature 
of the invisible decay of $Z'$ produced in the reaction  $e^- + Z \to e^- + Z + Z'; ~ Z' \to invisible$  due to mixing of Eq.(\ref{mixstr})
 is  the same - it is an event with a large missing energy carried away by the $Z'$. Thus, by using Eq.(\ref{mixstr})  and  bounds on the $\gamma - A'$ mixing 
 the NA64 can also set constraints on   coupling  $e_\mu$.
 
 The current  NA64 bounds on $\epsilon$ parameter for the  dark photon mass  region  $1\lesssim m_{Z'} \lesssim 10$ MeV are in the range 
 $ 10^{-5} \lesssim  \epsilon \lesssim 10^{-4}$ for the number of   accumulated electrons on target (EOT) $n_{EOT} \simeq 4.3 \cdot 10^{10}$ \cite{na64-2}. 
  Taking into account that the sensitivity of the experiment   scales as $\epsilon \sim 1/\sqrt{n_{EOT}}$,   
 results in  required  increase of statistics by a factor $\simeq$100  in order  to improve  sensitivity up to the  mixing value of 
 Eg.(\ref{loop}) for this $Z'$ mass   region. This would allow  either to discover  the  
$Z'$ or exclude it as an explanation of the $g_{\mu} -2 $ anomaly for the  substantial part of the  mass range $m_{Z'} \ll m_{\mu}$  by  using the electron beam. 
The direct search for the $Z'$ in missing-energy  events in the reaction  $\mu Z \rightarrow \mu Z Z'; Z'\to invisible$  in the dedicated experiment of Ref.\cite{gkm}  with the muon beam at CERN  would  then be an important cross check  of results obtained with the  electron beam. 
Let us note that the mixing given by the Eq.(6)  would also lead to an extra contribution to the elastic $\nu e \to \nu e$  scattering signal 
 in the solar neutrino measurement at the Borexino experiment \cite{takeshi1}.
 The BOREXINO data  on the elastic $\nu_{\mu}e$ scattering \cite{BOREXINO} lead to lower bound 
on $m_{Z^`} \geq (5-10)~MeV$ by assuming that muon anomaly is explained due to 
existence of light $Z^`$ boson interacting with $L_{\mu} - L_{\tau}$ current and 
there is no tree level mixing between photon and $Z^`$, i.e.  $\epsilon_{tree} = 0$.
 The measurement of $\nu - e$  elastic scattering  in the LSND experiment \cite{lsnd} set  a similar bound to the $e_\mu$ coupling for $m_{Z'} \lesssim 10$ MeV \cite{takeshi1}.
The expected 90\% C.L.  NA64 exclusion regions   in the ($m_{Z'}, e_\mu$) 
plane (dashed curves) from the measurements with the election beam for  $\simeq 4\times 10^{12}$ and   $\simeq 4\times 10^{13}$ EOT \cite{na64-1,na64-2,NA64exp}  and muon beams for 
$\simeq 10^{12}$  muons on target (MOT)  \cite{gkm} 
are shown in Fig.1. Constraints from the BOREXINO  \cite{takeshi1}, CCFR   \cite{ccfr}, and BABAR \cite{BABAR1} experiments,  as well as the BBN excluded area \cite{ takeshi1, kamada}  are also shown.
The parameter space shown in Fig.1 could  also be probed by other electron experiments such as Belle II \cite{takeshi2},  BDX \cite{bdx},  and LDMX \cite{ldmx},  which  would provide  important complementary results. 
 \begin{figure}[tbh!!]
\begin{center}
\includegraphics[width=0.8\textwidth,height=0.6\textwidth]{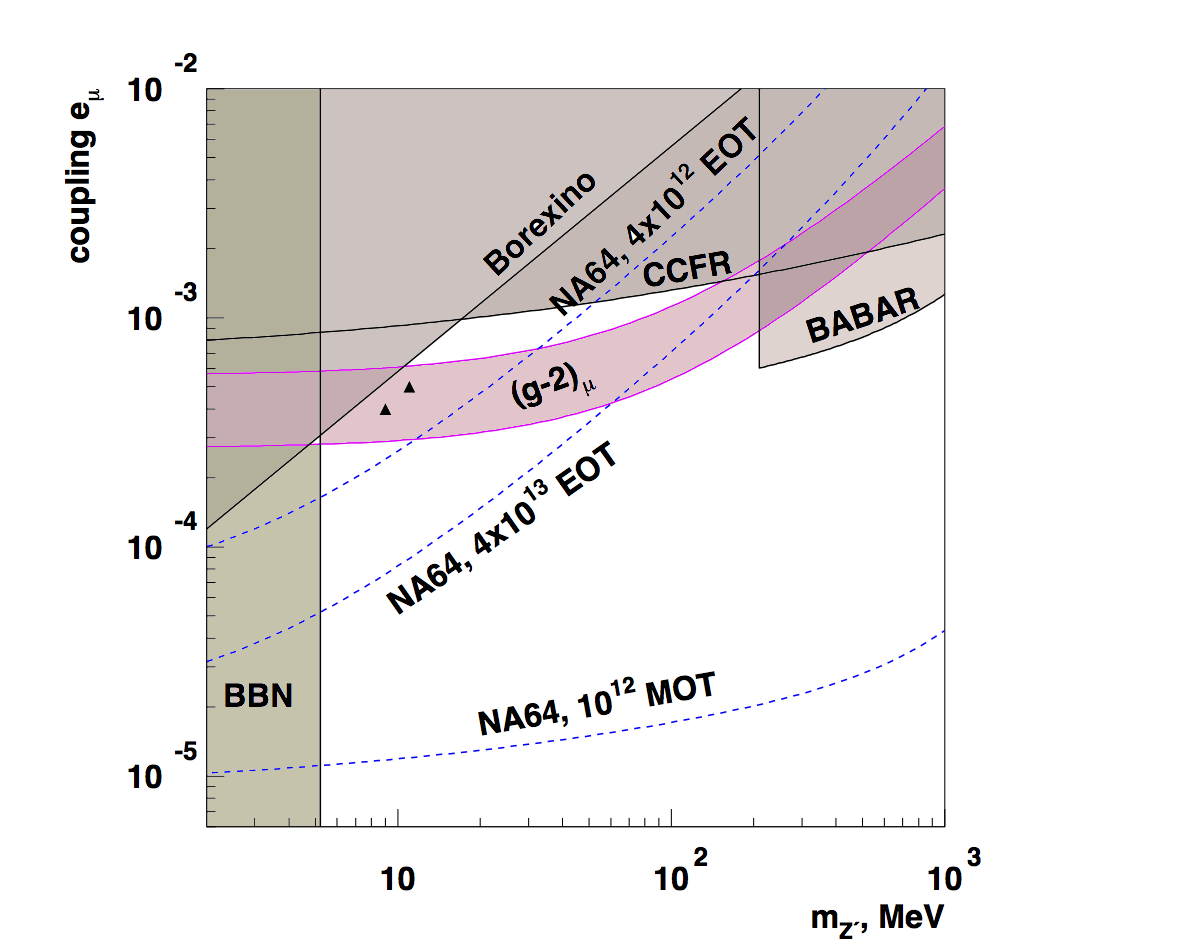}
\caption {The NA64 90\% C.L. expected exclusion regions in the ($m_{Z'}, e_\mu$) 
plane (dashed curves) from the measurements with the election (NA64, $\simeq 4\times 10^{12}$ EOT and $\simeq 4\times 10^{13}$ EOT) \cite{na64-1,na64-2,NA64exp}  and  muon (NA64-$\mu$, $\simeq 10^{12}$ MOT) \cite{gkm} beams.  
 Constraints from the BOREXINO  \cite{takeshi1}, CCFR   \cite{ccfr}, and BABAR \cite{BABAR1} experiments,  as well as the BBN excluded area \cite{ takeshi1, kamada}
 are also shown. 
  Two triangles indicate   reference points  corresponding to 
 the mass $m_{Z'} = 9$ and 11 MeV, and coupling $e_{\mu} = 4\times10^{-4}$ and $5\times10^{-4}$, respectively, 
 which are used to explain the IceCube results, see Ref.\cite{takeshi1} for details.
  \label{exclinv}}
\end{center}
\end{figure} 
\par Another  possible way to search for the $Z'$ is based on production and detection  of its visible decay mode, $Z' \to e^+ e^-$, 
which can also occur  at the one-loop  level.  The flux of $Z'$s  would be  generated  in a high
intensity beam dump experiment  through the mixing with photon produced either directly in the dump \cite{blum}  or,  e.g.,  in the $\pi^0, ~\eta,~\eta'$ decays \cite{sngdec}. The $Z'$s  would then penetrate  the dump without significant interaction and   decay in flight  into $e^+e^-$ pairs  which can be observed in a far detector. 
For a given flux $d\Phi(m_{Z'}, E_{Z'}, N_{POT})/d E_{Z'}$ of $Z'$'s from the dump 
the expected number of  $Z'\to \ee$ decays occurring within the fiducial length $L$ of a far  
detector located  at a distance $L'$ from the beam dump  is given by 
\begin{eqnarray}
N_{\zdecay}=  Br(Z' \to \ee) \int \frac{d\Phi}{dE_{Z'}} exp\Bigl(-\frac{L'm_{Z'}}{P_{Z'}\tau_{Z'}}\Bigr) \nonumber \\  
\cdot  \Bigl[1-exp\Bigl(-\frac{L m_{Z'}}{P_{Z'}\tau_{Z'}}\Bigr)\Bigr]\epsilon_{eff} A dE_{Z'}
\label{flux}
\end{eqnarray}
where $ E_{Z'}, P_{Z'}$, and $\tau_{Z'}$ are the $Z'$ energy, momentum and the lifetime  
at rest, respectively,  $\epsilon_{eff}, ~A $ are  the $\ee$ pair reconstruction efficiency and acceptance,
$N_{POT}$ is the  number of primary particles on target (dump).
For the mass region $1 \lesssim m_{Z'} \lesssim 200$ MeV the branching fraction is given by 
\begin{equation}
Br(Z' \to \ee) = \frac{\Gamma (\zdecay)} {\Gamma(\zdecay) + \Gamma (\znudecay)}  
\label{bratio}
\end{equation}
where the decay rate of the $Z'$ into neutrino, $\Gamma(\znudecay)$ ( $\nu = \nu_\mu, ~\nu_\tau$)  and $\ee$ pairs,  $\Gamma(\zdecay)$ 
is given by 
\begin{equation}
\Gamma(\znudecay) = \frac{e_\mu^2}{12\pi} m_{Z'} 
\label{znunu}
\end{equation}   
and 
\begin{eqnarray}
\Gamma (\zdecay) = \frac{\alpha}{3} \epsilon^2 m_{Z'}\sqrt{1-\frac{4m_e^2}{m_{Z'}^2}} \Bigl( 1+ \frac{2m_e^2}{m_{Z'}^2}\Bigr), 
\label{zee}
\end{eqnarray}
respectively.
Using Eqs.(\ref{bratio} - \ref{zee}) we found the $Z'$ lifetime and branching fraction  to be in the range $10^{-15} \lesssim \tau_{Z'} \lesssim 10^{-13}$ s. 
 This results in a very short $Z'$ decay  length $c \tau_{Z'} \gamma \simeq 150$ cm even for the $m_{Z'} \simeq 1$ MeV and  $E_{Z'} \simeq 50$ GeV.
Thus,  the attenuation of the $Z'$  flux due to $Z'$ decays in flight  which is given by the term $ exp\Bigl(-\frac{L'm_{Z'}}{P_{Z'}\tau_{Z'}}\Bigr)$ in Eq.(\ref{flux}), 
give a suppression factor $\ll 10^{-15}$ for any beam dump experiment searching for an excess of $\ee$ pairs from dark photon decays \cite{pdg}, 
as they typically used $L'\gtrsim 100$ m and  $Z'$ energy range $E_{Z'} \lesssim 50$ GeV.  
Because the effective coupling of $Z_{\mu}$ to electrons (or quarks) due to the mixing of  Eq.(\ref{mixstr})  is suppressed by a factor $ \approx 10^{-2}$,  the branching fraction $Br(Z' \rightarrow e^+e^-) $  is  estimated to be  $\simeq O(10^{-4})$.
Taking all these into account makes current constraints $10^{-8} \lesssim \epsilon \lesssim 10^{-4}$ \cite{pdg}  from the beam dump experiments  searching for visible $A' \to \ee$ decays of dark photons in the mass range $1 \lesssim m_{A'} \lesssim 200$ MeV inapplicable to the $Z'$ case and much more weaker than the value of 
Eq.(\ref{loop})  as they  were obtained under the assumption that this decay mode is  dominant.   
\par Thus, the  advantage of searching for $Z'$ in a  missing-energy type experiment, e.g. such as  NA64,   is that its sensitivity is roughly proportional
to the mixing squared, $\epsilon^2$  associated with the $Z'$ production
in the primary reaction and its subsequent invisible
decay, while for the visible case it is proportional to $\epsilon^2 \times Br(Z' \to e^+e^-)$.  The  factor 
$\epsilon^2$ is  coming from the $Z'$ production process and
another suppression factor $Br(Z' \to e^+e^-) =  O(10^{-4})$  from the $Z' \to e^+ e^-$ decay  in the  detector.
Similar arguments are also valid for the experiments that searched for the $A'$ in particle decays, because their exclusion area  is  
 $\epsilon \gtrsim 10^{-4}-10^{-3}$  for the mass range $1 \lesssim m_{A'} \lesssim 200$ MeV \cite{pdg}. 
 As a consequence, taking into account  the previous discussions, in any beam dump or decay experiment 
using electrons or quarks as a source of $Z'$s through the mixing of (6), the number of visible  
$Z' \to e^+ e^-$ signal events would be highly suppressed resulting in a weak  bound on $\alpha_{\mu}$ . 
 \par Similar considerations  results in rather modest   constraints  on  invisible decays of $Z'$ which one can extract  from the present results of  dark-photon experiments searching for the invisible $A'$ decays \cite{pdg}. 
For example, the bound on the coupling $\alpha_\mu$ from the  $K^+\to \pi^+ + missing~energy$ decay is at the level $\alpha_{\mu} \leq O(10^{-3})$, which is several orders of magnitude below the value from Eq.(4).
\par  Finally, note that in order to cover the range $\epsilon \lesssim 10^{-5}$ for the $Z' \to e^+ e^-$ decays the 
trick would  be to try to run a  corresponding experiment  in a very short-length  beam dump mode. A good  example of such approach is the AWAKE experiment, which 
plan to search for dark photon decays $A' \to \ee$  with a $\simeq 50$ GeV electron beam by using short W-dump and a
detector  located at a distance $L' \simeq$ a few meter  \cite{awake}. This experiment would be very complementary to the $Z'$ searches in invisible decay  
mode provided the accumulation of $\gtrsim 10^{16}$ EOT is feasible.  Another experiment,  which potentially might be sensitive to the values around of those from of Eqs.(\ref{mixstr}),(5) for the masses  $ m_{Z'} \simeq 100$ MeV,  
is the HPS \cite{hps}, which currently  aims at reaching the sensitivity $\epsilon \lesssim 10^{-5}$ for the   $A'\to \ee$ decays.  
\par Let us  show now that an extension of the $L_{\mu} - L_{\tau}$ model is able to explain today DM density in the Universe. 
 Consider the simplest SM extension with an additional complex scalar 
field $\phi_d$\footnote{The annihilation cross-section for scalar DM 
has $p$-wave suppression that allows to escape CMB bound \cite{Planck}.}.  
The charged dark 
matter field  $\phi_d$ interaction with   the $Z'$ field is 
\begin{equation}
L_{\phi Z'} = (\partial^{\mu}\phi - ie_dZ'^{\mu}\phi)^{*}(\partial_{\mu}\phi - 
ie_dZ'_{\mu}\phi) - m^2_{DM}\phi^{*}\phi -  \lambda_{\phi}  (\phi^{*} \phi )^2  \,
\label{lagrangian1} 
\end{equation} 
The annihilation cross section  $\phi_d \bar{\phi_d} \rightarrow \nu_{\mu}\bar{\nu}_{\mu}, \nu_{\tau}\bar{\nu}_{\tau}$  
 for $s \approx 4 m^2_{DM}$
has the form\footnote{Here we consider the case $m_{Z'} > 2 m_{DM}$.}
\begin{equation}
\sigma v_{rel} = \frac{8\pi}{3} \frac{\epsilon^2\alpha\alpha_d m^2_{DM}v^2_{rel}}
{(m^2_{Z'} - 4 m^2_{DM})^2 } \,,
\label{crsec}
\end{equation} 
We use standard assumption  that in the hot early Universe DM is in equilibrium with 
ordinary matter. During the Universe expansion the temperature decreases and at some 
point the thermal decoupling of the Dark Matter starts to work. Namely, at 
some freeze-out temperature the  cross-section of annihilation $DM~particles\rightarrow ~SM~ particles$ 
 becomes too small to obey the equilibrium of DM particles with 
the SM particles and DM decouples. The experimental data are in favour of scenario          
with cold relic for which the freeze-out temperature is much lower 
than the mass of the particle. In other words DM particles decouple in 
the non-relativistic regime. The value of the DM annihilation 
cross-section at the decoupling epoch determines the value of 
the current DM density in the Universe. Too big annihilation cross-section 
leads to small DM density and vise versa too small annihilation 
cross section leads to DM overproduction. The observed 
value of the DM density $\frac{\rho_{DM}}{\rho_{c}} \approx 0.23$ allows 
to estimate the  DM annihilation cross-section 
into the SM particles and hence to estimate the discovery potential of light dark 
matter both in direct underground and accelerator experiments. 
\par The dark matter relic density can be numerically estimated as \cite{39}
\begin{equation}
\Omega_{DM}h^2 = 0.1\Bigl[\frac{(n+1)x_f^{n+1}}{(g_{*s}/g^{1/2}_*)}\Bigr]\frac{0.856\cdot 10^{-9}GeV^{-2}}{\sigma_0} \,,
\end{equation}
where $<\sigma v_{rel}> = \sigma_o x^{-n}_f$, $x_f = \frac{m_{DM}}{T_{dec}}$ 
   and 
\begin{equation}
x_f = c - (n + \frac{1}{2}){\rm ln}(c) \,,
\end{equation}
\begin{equation}
c = {\rm ln}\Bigl[0.038(n+1)\frac{g}{\sqrt{g_*}}M_{Pl}m_{DM}\sigma_0\Bigr]\\,.
\end{equation}

For the case where dark matter consists of dark matter particles and dark matter antiparticles 
the $DM \bar{DM} \rightarrow SM~ particles$ annihilation cross sestion $\sigma = \frac{\sigma_{an}}{2}$.
Numerically we find that 
\begin{equation}
 k(m_{DM}) \cdot 10^{-6}\cdot(\frac{m_{DM}}{GeV})^2 \cdot\Bigl[\frac{m^2_{A^`}}{m^2_{DM}} - 4\Bigr]^2 
 = \epsilon^2 \alpha_{D} \,.
\end{equation}
Here the coefficient $k(m_{DM})$ depends logarithmically on the dark matter mass $m_{DM}$ and 
$k_{DM} \approx 0.5(0.9)$ for $m_{DM} =1(100)~MeV$.
For instance, for $m_{A'} = 2.2~m_{DM}$ we have
\begin{equation}
	0.71 k(m_{DM}) \cdot  10^{-6}\cdot\Bigl[\frac{m_{DM}}{1~{\rm GeV}}\Bigr]^2 = \epsilon^2 \alpha_{d} \,.
\end{equation}
As a consequence of (14) we find that 
for $m_{Z'} \ll m_{\mu}$ the values   $\epsilon^2 = (2.5 \pm 0.7) \cdot 10^{-6} $ and 
\begin{equation}
\alpha_{d} =
(0.28  \pm 0.08)k(m_{DM})  \cdot \Bigl[\frac{m_{DM}}{1~{\rm GeV}}\Bigr]^2  
\end{equation}
explain both 
the  $g_{\mu} - 2$ muon anomaly and today DM density. 
We can rewrite the equation (15) in the form 
\begin{equation}
\frac{e_d^2}{e^2_{\mu}} = 
(16  \pm 9) k(m_{DM}) \cdot \Bigl[\frac{m_{DM}}{1~{\rm MeV}}\Bigr]^2  \,.
\end{equation}
So we see that for $m_{DM} \geq 1$~MeV we have $e_d \gg e_{\mu}$, i.e.the  $Z'$ must interact much more strongly with light DM than with the SM matter. 

 In summary,   the $L_\mu - L_\tau$ model with the light vector boson $Z'$ interacting with  $L_{\mu} - L_{\tau}$ 
current   is a well-motivated SM extension, with impressive
indirect support from the possible explanation of the muon $g_{\mu} -2$ anomaly and  several observations in  neutrino sector and astrophysics. 
While the model can be effectively tested with the direct high-energy muon experiment at the CERN SPS \cite{gkm}, we show that  
  nonzero $\gamma-Z'$ mixing  generated in the model at the one-loop level strongly motivates  the complementary searches of the light  $Z'$ 
  with high-energy electron beams. 
  This open up an intriguing possibility for  probing  the $L_{\mu} - L_{\tau}$ gauge boson  $Z'$    in the  near future with ongoing NA64 experiment with the  statistics  increased by a factor $\simeq10 -100$.  The $Z'$ searches can be as well performed in the incoming   dark photon  experiments, e.g such as  AWAKE,  Belle-II,  BDX,  and LDMX.  Moreover an extension of the $L_{\mu} - L_{\tau}$     model allows to explain relic Dark Matter density
for $m_{Z'} \simeq O(10)$ MeV, which  strengthen motivation for the  experimental search of  the 
 $L_{\mu} - L_{\tau}$ mediator of the DM production in invisible decay mode.  
 Finally, we note that if the  $Z'$ couples to light DM, then an
additional contribution from the invisible  decay mode   $Z' \rightarrow  dark~ matter$   
increases the $Z'\rightarrow  invisible$ decay rate as a consequence  for $m_{Z^`} > 2m_{\mu}$ visible decay $Z^` \rightarrow  \mu^+ \mu^- $ 
is suppressed.

This work grew in part from our participation in the 2nd Annual Physics Beyond Colliders workshop. We wish to thank organizers of this conference for their warm hospitality at CERN. We thank members of the PBC BSM working group, in particular G. Lanfranchi, J. Jaeckel,  and A. Rozanov, for discussions  and  valuable comments.  We are indebted to Prof. V.A. Matveev and  our colleagues from 
the NA64 and AWAKE Collaborations  for many for useful suggestions. 

\end{document}